\documentclass[twocolumn,showpacs,pra]{revtex4}

\usepackage{amsmath, amsthm, amssymb}
\usepackage{graphicx}
\usepackage{dcolumn}
\usepackage{bm}

\newcommand{\de}[1]{\left( #1 \right)}

\newcommand{\DE}[1]{\left\{ #1 \right\}}

\renewcommand{\Re}[1]{{\mathrm{Re}}\de{#1}}

\newcommand{\ket}[1]{\left| #1 \right\rangle}
\newcommand{\bra}[1]{\left\langle #1 \right|}

\newcommand{\norm}[1]{\left\| #1 \right\|}

\newcommand{\tr}{\mathrm{Tr}}

\newcommand{\ie}{{\it{i.e.~}}}

\newtheorem{definition}{Definition}

\begin{document}

\title{Multipartite entanglement of superpositions}

\author{D. Cavalcanti}\email{Daniel.Cavalcanti@icfo.es}
\affiliation{ICFO-Institut de Ciencies Fotoniques, Mediterranean
Technology Park, 08860 Castelldefels (Barcelona), Spain}

\author{M. O. Terra Cunha}\email{tcunha@mat.ufmg.br}
\affiliation{Departamento de Matem\'atica, Universidade
Federal de Minas Gerais, Caixa Postal 702, 30123-970, Belo
Horizonte, MG, Brazil}
\affiliation{School of Physics and Astronomy, University of Leeds, Leeds, LS2 9JT, UK}

\author{A. Ac\'in}\email{Antonio.Acin@icfo.es}
\affiliation{ICFO-Institut de Ciencies Fotoniques, Mediterranean
Technology Park, 08860 Castelldefels (Barcelona), Spain}
\affiliation{ICREA-Instituci\'o Catalana de Recerca i Estudis Avan\c
cats, Lluis Companys 23, 08010 Barcelona, Spain}

\begin{abstract}
The \emph{entanglement of superpositions} [Phys. Rev. Lett. {\bf
97}, 100502 (2006)] is generalized to the multipartite scenario:
an upper bound to the multipartite entanglement of a superposition
is given in terms of the entanglement of the superposed states and
the superposition coefficients. This bound is proven to be tight
for a class of states composed by an arbitrary number of qubits.
We also extend the result to a large family of quantifiers which
includes the negativity, the robustness of entanglement, and the
best separable approximation measure.
\end{abstract}

\pacs{03.67.-a, 03.67.Mn}

 \maketitle

\section{Introduction}

The study of quantum correlations is certainly one of the most
challenging issues that physicists have been faced with. Both from
an experimental and a theoretical point of view the
characterization of entanglement has proven to be very hard
\cite{HorReview}. Even in the simplest scenario, namely the study
of bipartite pure-state entanglement, we still find open
questions. Needless to say, the case of mixed states and
multipartite systems are much richer and further to be completely
understood.

In a recent work \cite{LPS}, Linden, Popescu and Smolin have
raised the following question: Given pure states $\ket{\Psi}$ and
$\ket{\Phi}$ on a bipartite system, how is the entanglement of the
superposition state
\begin{equation}\label{state}
\ket{\Gamma}=a\ket{\Psi}+b\ket{\Phi},
\end{equation}
related to the entanglement of the constituents $\ket{\Psi}$ and
$\ket{\Phi}$ and to the coefficients $a$ and $b$? This apparently
simple question was shown to exhibit a rich answer in terms of
nontrivial inequalities relating these quantities. In order to
quantify the entanglement, the authors of \cite{LPS} used the Von
Neuman entropy of the reduced state (often called Entanglement
Entropy \cite{Benn1}). This is a natural choice since this
quantifier has a clear operational meaning: it gives the number of
Bell pairs that can be produced from a large number of copies of
an arbitrary entangled state by local operations and classical
communication \cite{Benn1}. However other entanglement quantifiers
can also be used and, in fact, distinct bounds for the
entanglement of a superposition can be find depending on this
choice ~\cite{YYS,Ou}.

The main goal of this work is to generalize the ideas raised in
\cite{LPS} to the multipartite scenario. However, instead of
working with a specific entanglement quantifier, we have chosen a
family of quantifiers called \emph{witnessed entanglement}
\cite{Bra}. This family represents those measures that can be
written as
\begin{equation}\label{Ew}
E_{\mathcal{W}}(\rho)= \max \{0,-\min_{W \in {\mathcal{W}}}\tr(W\rho)\},
\end{equation}
where ${\mathcal{W}}$ is a restricted set of entanglement witnesses
\cite{Wit}. The term entanglement witness refers to a Hermitian
non-positive operator which has positive mean value for all
separable states, hence a negative mean value indicates the presence
of entanglement \cite{Wit,Terh}. For an entangled pure state $\rho =
\ket{\psi}\bra{\psi}$, the witnessed entanglement can be expressed
by \cite{note1}
\begin{equation}
E_{\mathcal{W}}(\psi)=-\bra{\psi}W^\psi_{opt} \ket{\psi},
\end{equation}
being $W^\psi_{opt}$ an optimal witness for the state $\ket{\psi}$
(\ie a witness satisfying the minimization problem in \eqref{Ew}).
This simplified way of writing $E_{\mathcal{W}}$ will be particularly useful
to our constructions.

One important fact concerning $E_{\mathcal{W}}$ is that several interesting
entanglement quantifiers belong to this class. These quantifiers
include concurrence \cite{Con}, negativity \cite{Neg1,Neg2},
robustness of entanglement \cite{Rob1,Rob2,GenRan}, and the best
separable approximation \cite{BSA}. Each one of these examples can
be written in the form of Eq. \eqref{Ew} by changing the choice of the
set ${\mathcal{W}}$ \cite{Bra}. Another advantage of $E_{\mathcal{W}}$ is that
it can be directly linked to measurable quantities since $W$ is a
Hermitian operator. Because of that, $E_{\mathcal{W}}$ can be experimentally
estimated even for an unknown quantum state  \cite{CavTerra,Eis}.
It must be stressed, and this is very important in our considered
scenario, that $E_{\mathcal{W}}$ can also quantify different kinds of
multipartite entanglement: the restricted set ${\mathcal{W}}$ can be chosen as
a set of entanglement witnesses which detect only a certain kind
of entanglement. Besides that, among the witnessed entanglement
quantifiers, we can find both operational measures
\cite{Neg3,Bra2,Mar} (\ie entanglement quantifiers with some
operational meaning) and geometrical ones \cite{BSA,Ber,Cav} (\ie
quantifiers related to geometrical aspects of the state space).

\section{Multipartite entanglement of superpositions}

The main scope of this work is to obtain an upper bound to the
witnessed entanglement of the state \eqref{state} based on the
entanglement of the superposed states $\ket{\Psi}$ and
$\ket{\Phi}$ and the coefficients appearing in the superposition. In this section, we first derive an inequality
relating these quantities and then prove its tightness. The witnessed entanglement of $\ket{\Gamma}$ can be written as
\begin{widetext}
\begin{eqnarray}
E_{\mathcal{W}}(\Gamma)&=&\max\{0,-\min_{W \in
{\mathcal{W}}}\bra{\Gamma}W\ket{\Gamma}\}\nonumber\\&=&\max\{0,-\min_{W \in
{\mathcal{W}}}[|a|^2\bra{\Psi}W\ket{\Psi}+|b|^2\bra{\Phi}W\ket{\Phi}+2\Re{ a^*
b\bra{\Psi}W\ket{\Phi}}]\},
\end{eqnarray}
an expression that resembles the usual interference pattern
originated by superpositions. For finite dimension the minimization
problem is solved using the so-called optimal entanglement witness
$W_{opt}$ (inside the set ${\mathcal{W}}$ which defines the quantifier). So we
can write
\begin{equation}
E_{\mathcal{W}}(\Gamma) =
\max\{0,-|a|^2\bra{\Psi}W_{opt}^{\Gamma}\ket{\Psi}-|b|^2\bra{\Phi}W_{opt}^{\Gamma}\ket{\Phi}-2\Re{
a^*b\bra{\Psi}W_{opt}^{\Gamma}\ket{\Phi}}\}.
\end{equation}
Again, $W_{opt}^{\Gamma}$ denotes a witness that is optimal for
the state $\ket{\Gamma}$. Different states usually have different
optimal entanglement witnesses. We are naturally led to the
inequality
\begin{eqnarray}
E_{\mathcal{W}}\de{\Gamma} &\leq& \max\{0,-|a|^2\bra{\Psi}W_{opt}^{\Psi}\ket{\Psi}\}
+\max\{0,-|b|^2\bra{\Phi}W_{opt}^{\Phi}\ket{\Phi}\} + \max\{0,-2\Re{a^*
b\bra{\Psi}W_{opt}^{\Gamma}\ket{\Phi}}\}\nonumber\\
&=& |a|^2 E_W(\Psi)+|b|^2 E_W(\Phi)+2\max\{0,-\Re{a^*
b\bra{\Psi}W_{opt}^{\Gamma}\ket{\Phi}}\},
\end{eqnarray}
\end{widetext}
where we have also made use of the inequality
$\max\{0,a+b\}\leq\max\{0,a\}+\max\{0,b\}$. Attention must now be
payed to the interference term. The Cauchy-Schwarz inequality
implies
\begin{equation}
\label{Wquote} E_W\de{\Gamma}\leq |a|^2 E_W\de{\Psi}+|b|^2
E_W\de{\Phi} + 2|a||b|\norm{W_{opt}^{\Gamma}}.
\end{equation}
Note that the normalization of the kets involved was used and we
take the norm of an operator as its maximal singular value.
Expression \eqref{Wquote} relates the entanglement of
$\ket{\Gamma}$ to the entanglement of each one of the superposed
states (and the coefficients of the superposition) but also
depends on the form of the optimal entanglement witness
$W_{opt}^{\Gamma}$. This dependence on the optimal entanglement
witness is expected as the
restrictions in $W_{opt}^{\Gamma}$ imply the features of the
entanglement quantifier we are dealing with.

At this point it is worth asking if inequality \eqref{Wquote} can
be saturated. Let us choose the negativity as a quantifier for
instance. In this case we can compute $W_{opt}^{\Gamma}$
analytically. For a given state $\rho$, it is given by the partial
transposition of the projector onto the subspace of negative
eigenvalues of $\rho^{T_A}$, where $\rho^{T_A}$ denotes the
partial transposition of $\rho$ \cite{optW}. It is now easy to see
that for the two-qubit states $\ket{\Phi}=\ket{00}$ and
$\ket{\Psi}=\ket{11}$, the inequality \eqref{Wquote} becomes
$|a||b|\leq|a||b|$.

In the previous examples we used the fact that we knew the optimal
witness $W_{opt}^{\Gamma}$. Let us now remove this strong
assumption. It was shown in Ref. \cite{Bra} that if ${\mathcal{W}}$ (in
Eq.~\eqref{Ew}) is the set of entanglement witnesses satisfying
$-nI\leq W\leq mI$, where $m,n\geq 0$, $E_{\mathcal{W}}$ is an entanglement
monotone \cite{EntMon}. Setting
$k=\max(m,n)$ we have
\begin{equation}\label{GenIneq}
E_{\mathcal{W}}\de{\Gamma}\leq |a|^2 E_{\mathcal{W}}(\Psi)+|b|^2 E_{\mathcal{W}}(\Phi) + 2k|a||b|.
\end{equation}

As our main goal here is to work in the multipartite case it would
be interesting to find examples of multipartite states for which
relation \eqref{GenIneq} is saturated. The main barrier to be
overcome in this case is the fact that it is not known, in
general, how to compute multipartite entanglement quantifiers.
Nevertheless we develop a way of calculating the generalized
robustness of entanglement for GHZ-like states and use this
information to prove the tightness of inequality \eqref{GenIneq}
regardless the number of particles involved.

The generalized robustness of entanglement \cite{GenRan} admits two
representations, one in terms of how robust the entanglement of a
state is against arbitrary noise and the other as a witnessed
entanglement. Let us present both definitions precisely.
\begin{definition}
The generalized robustness of entanglement of a
state $\rho$ is given by
\begin{equation}
R_g(\rho)=\inf_{\pi \in \mathcal{D}}\min\DE{s : \sigma \de{\rho,\pi,s} \in \mathcal{S}}, 
\end{equation}
where $\sigma$ denotes the state
\begin{equation}\label{sigma}
\sigma\de{\rho,\pi,s}=\frac{\rho+s\pi}{1+s},
\end{equation}
$\mathcal{D}$ the set of all density operators, and $\mathcal{S}$ the set of
 separable ones (with respect to the specific form of entanglement that is considered).
\end{definition}
\begin{definition}
$R_g(\rho)$ is the witnessed entanglement $E_{\mathcal{W}}
\de{\rho}$ when ${\mathcal{W}}$ is the set
of witness operators satisfying $W \leq I$.
\end{definition}
The equivalence of these definitions was proven in \cite{Bra}. We
make use of both to show that for the $N$-qubit family of states
\begin{equation}\label{family}
\ket{GHZ_N(\phi)}=\frac{\ket{0^{\otimes
^N}}+e^{i\phi}\ket{1^{\otimes ^N}}}{\sqrt{2}}, \end{equation} the
inequality \eqref{GenIneq} is saturated. Clearly if one chooses an
arbitrary state $\pi$ such that the state $\sigma\de{\rho,\pi,s}$ is
separable for some value of $s$, this number $s$ gives an upper
bound for the value of $R_g(\rho)$. On the other hand, taking an
arbitrary entanglement witness $W$ for the state $\rho$ satisfying
the condition $W<I$ , $-\tr(W\rho)$ gives a lower bound to
$R_g(\rho)$ according to \eqref{Ew}. We will now establish lower
and upper bounds for $R_g(GHZ_N(\phi))$ that turn out to be equal,
getting the exact value of this quantity and also the value of $k$
needed for the bound \eqref{GenIneq}.\newline
\emph{Upper bound.} Consider, in Eq. \eqref{sigma},
\begin{equation}
\rho=\ket{GHZ_N(\phi)}\bra{GHZ_N(\phi)}
\end{equation}
and
\begin{equation}
\pi=\ket{GHZ_N(\phi)_{\bot}}\bra{GHZ_N(\phi)_{\bot}},
\end{equation}
where
\begin{equation}
\ket{GHZ_N(\phi)_{\bot}}=\frac{\ket{0^{\otimes
^N}}-e^{i\phi}\ket{1^{\otimes ^N}}}{\sqrt{2}}.
\end{equation}
Using the Peres criterion \cite{PerHor} we see that $\sigma$ has
positive partial transposition only for $s=1$. Moreover, for this
point it can be directly verified that $\sigma$ is also separable.
So we get
\begin{equation}\label{upper}
R_g(GHZ_N(\phi))\leq1.
\end{equation}
\emph{Lower bound.} The following operator is a genuine
entanglement witness for the state $\ket{GHZ_N(\phi)}$
\cite{Wei,Cav}:
\begin{equation}\label{W}
W=I-2\ket{GHZ_N(\phi)}\bra{GHZ_N(\phi)},
\end{equation}
which clearly satisfies the condition $W<I$. Hence,
 definition \eqref{Ew} leads to
\begin{equation}\label{lower}
-\tr(W\ket{GHZ_N(\phi)}\bra{GHZ_N(\phi)})=1\leq R_g(GHZ_N(\phi)).
\end{equation}
As the upper bound \eqref{upper} and lower bound \eqref{lower}
coincide we have that $R_g(GHZ_N(\phi))=1$, and can also conclude
that the witness \eqref{W} satisfies the minimization problem in
\eqref{Ew}. It then allows us to extract the value $k=1$.

Putting all these facts together we conclude that the inequality
\eqref{GenIneq} saturates for the class of states \eqref{family}.

\section{Conclusions}

We extended the notion of \emph{entanglement of superpositions} to
the multipartite scenario. An inequality relating the entanglement
of quantum states to the entanglement of the state constructed
through their superposition was found. This inequality was proven to
be tight for a family of $N$-qubit states and a choice of
entanglement quantifier. Moreover a large class of entanglement
quantifiers, with both operational and geometrical meanings, was put
in this context.

It is also worth to note that the inequalities derived here can be
extended for the case where more than two states are superposed
\cite{xiang}. Future research could include the study of other
examples of states and quantifiers treated in our general
perspective.

\section{acknowledgements}
The authors thank F. Brand\~ao for comments on this manuscript. D.
Cavalcanti acknowledges hospitality in the University of Leeds
where part of this work was done. M.O. Terra Cunha thanks ICFO for
hospitality, and also CNPq, PRPq-UFMG, and Fapemig for funding.
This work is supported by Brazilian Millennium Project on Quantum
Information, the EU Qubit Applications Project (QAP) Contract
number 015848, the Spanish projects FIS2004-05639-C02-02 and
Consolider QOIT, and the Generalitat de Catalunya.


\begin{thebibliography}{30}
\bibitem{note1}
We suppose $\ket{\psi}$ to have the kind of entanglement
to which $W^{\psi}_{opt}$ is constructed to witness. Remember that
in the multipartite case a state can show different kinds of
entanglement, and possibly the set ${\mathcal{W}}$ is tailored to
witness one kind of entanglement, while $\ket{\psi}$ can show only
other kinds of entanglement.
\end{thebibliography}
\end{document}